\documentclass[aps,pre,twocolumn,showpacs]{revtex4}
\usepackage{epsfig}
\usepackage{amsmath,amssymb,amsthm}
\usepackage{graphicx}
\usepackage{psfrag}
\usepackage{bm}

\begin{document}
\title{Kardar-Parisi-Zhang interfaces bounded by long-ranged
potentials}

\author{Omar Al Hammal$^1$,  Francisco de los Santos$^1$, 
 Miguel A. Mu\~noz$^1$ \\ and Margarida M. Telo da Gama$^2$}

\affiliation{$^1$Departamento de Electromagnetismo y F{\'\i}sica de la
Materia and Instituto Carlos I de F{\'\i}sica Te\'orica y
Computacional, Universidad de Granada, Fuentenueva s/n, 18071 Granada,
Spain \\ $^2$Centro de F\'\i sica Te\'orica e Computacional e
Departamento de F\'\i sica da Faculdade de Ci\^encias da Universidade
de Lisboa, Avenida Professor Gama Pinto, 2, P-1649-003 Lisboa Codex, 
Portugal}


\begin{abstract}
We study unbinding transitions of a non-equilibrium
Kardar-Parisi-Zhang (KPZ) interface in the presence of long-ranged
substrates. Both attractive and repulsive substrates, as well as
positive and negative Kardar-Parisi-Zhang nonlinearities, are
considered, leading to four different physical situations. A detailed
comparison with equilibrium wetting transitions as well as with
non-equilibrium unbinding transitions in systems with short-ranged
forces is presented, yielding a comprehensive picture of unbinding
transitions, and of their classification into universality
classes. These non-equilibrium transitions may play a crucial role in
the dynamics of wetting or growth of systems with intrinsic
anisotropies.

\end{abstract}
\maketitle

\section{Introduction}    

Spatial constraints in systems where two (or more) bulk phases coexist
may lead to wetting transitions. This is the case, for example, of
confined fluids where one of two coexisting equilibrium phases (the
liquid, say) is in contact with a substrate with an interface
separating it from the second phase (the gas) at infinity. The liquid
{\em does not wet} the substrate if the thickness of the liquid film
is finite (there is a microscopic quantity of liquid). On the other
hand, the substrate is {\em wet} if there is a macroscopically thick
liquid film on it.  A wetting transition is said to occur when the
substrate changes from not being wet by the liquid to being wet.
Typically, two types of wetting transitions can be considered: by
increasing the temperature at bulk coexistence one may find either
{\it critical wetting} or a discontinuous transition; by varying the
chemical potential while the temperature is fixed above the wetting
transition temperature one finds a {\it complete wetting} transition,
at bulk coexistence. Under equilibrium conditions, a completely
analogous transition (often called drying) may occur when the
substrate preferentially adsorbs the gas phase \cite{reviews_wetting}.

Effective interfacial potentials are useful coarse-grained models that
have played a key role in understanding a large variety of equilibrium
wetting problems \cite{Zia,reviews_wetting}. These potentials, $V(h)$,
are functionals of the interfacial height (measured from the
substrate), $h({\bf x})$. In this framework, wetting transitions are
described as the unbinding of the (say liquid-vapor) interface from
the substrate, with the effective binding potential determined by the
microscopic forces between the constituents of the substrate and those
of the bulk phases. Typically, exponentials and power-law decaying
potentials $V(h)$ have been considered for systems dominated by
short-ranged and long-ranged forces, respectively.

There exist a large amount of phenomena describable in terms of
equilibrium wetting, either under short-range or under long-ranged
interactions, while it has only recently been recognized that
non-equilibrium effects, such as anisotropies in the interface growing
rules, may play a crucial role in describing some experimental
situations. Within this perspective, short-ranged {\it non-equilibrium
wetting} has been studied \cite{Haye-wetting,Lisboa}, and some
interesting novel phenomenology has been elucidated (see
\cite{reviews} for recent reviews). In particular, liquid-crystals
\cite{LQ}, molecular-beam epitaxial systems, as $GaAS$ \cite{GAAS}, or
materials exhibiting Stranski-Krastanov instabilities \cite{SK},
appear to be good candidates to require a non-equilibrium wetting
description. However, some of these systems, as well as many others
not enumerated, might include effective long-range substrate-interface
effects as also occurs in the equilibrium case.

Our goal in this paper is to fill this gap by providing a general and
systematic theory of non-equilibrium wetting under the presence of
effective long-ranged interactions. First, we briefly review the
equilibrium situation to set up the theoretical framework and,
afterwards, generalize it to embrace non-equilibrium situations.

In equilibrium, two types of analytical approaches are available:
static studies based on the ensemble theory
\cite{reviews_wetting} and dynamical, stochastic approaches that allow investigating
relaxational aspects. The second approach is amenable to
non-equilibrium extensions and is the one we employ. 
Thus, consider the simple Edwards-Wilkinson
dynamics \cite{HZ} subject to a bounding force (i.e. the derivative of
the bounding potential) \cite{Lipowsky85}:
\begin{equation}
\partial_t h({\bf x},t) = \nabla^2 h + a -\frac{\partial V(h)}{\partial h}+
\sigma \eta({\bf x},t). 
\label{EWwall}
\end{equation}
This includes ({\bf i}) the usual diffusion term, computed as minus the
derivative of a standard surface-tension term, 
({\bf ii}) the driving force, $a$, related to the chemical potential
difference between the two phases,
({\bf iii}) the Gaussian white
noise, $\eta({\bf x},t)$, and ({\bf iv}) the bounding force, which may derive 
from a short-ranged potential 
\begin{equation}
V(h)=  \frac{b}{p} ~e^{-p h} + \frac{c}{q} e^{-q h},
\label{SR}
\end{equation}
or from a long-ranged one
\begin{equation}
V(h)= \frac{b}{p h^{p }} + \frac{c}{q h^{q}},
\label{LR}
\end{equation}
where, $b$, $c>0$, and $p<q$ are parameters.  This last form,
Eq.(\ref{LR}), is known to be the correct functional form for systems
where the molecules interact through van der Waals forces
\cite{rusos}.

\begin{figure}
\centerline{\psfig{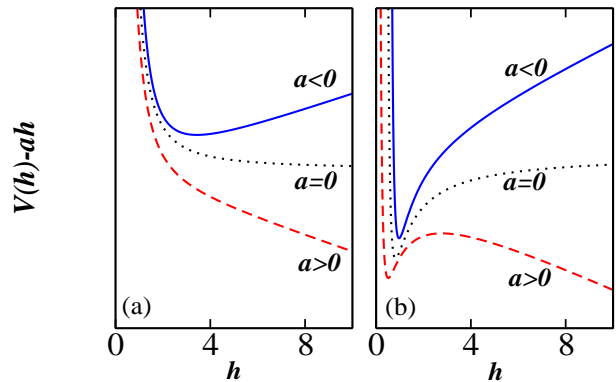}}
\caption{ Effective potentials as derived from Eq. (\ref{LR}) in
the pinned ($a<0$) and the depinned ($a>0$) phases, and at coexistence
($a=0$).  (a) repulsive walls ($b,c>0$); (b) attractive walls ($b<0$,
$c>0$).}
\label{fig1}
\end{figure}

By varying the chemical-potential, $a$, one controls the average
interfacial distance from the wall: small for $a < a_c$ (non-wet
phase), large for $a \approx a_c$, and increasing steadily with time
for $a > a_c$ (wet phase), i.e. the system exhibits an unbinding
transition at $a=a_c$.  The interface potentials $V(h)$ are, in all
cases, harshly repulsive at small $h$ to model the impenetrability of
the substrate. The parameter $b$ vanishes linearly with the
temperature, at the (mean-field) critical wetting temperature, and
represents the affinity or preference of the substrate for one of the
bulk phases (usually the liquid). We consider three distinct
situations (see Fig. \ref{fig1}):

\noindent{\it 1. Repulsive potential: complete wetting}.
If $b>0$ the 
potential describes the presence of a bounding substrate alone. In
this case, the broken symmetry induced by the substrate leads to the
divergence of the average position of the interface, at coexistence,
$a_c=0$: i.e. the system undergoes a {\it complete wetting}
transition.

The latter is described by Eq. (\ref{EWwall}) with the potential taken from
Eq. (\ref{SR}).
%
%
Two different regimes depending on the value of $p$ 
have been reported: for $p<2$ mean-field scaling holds
and $\langle h \rangle \sim t ^{1/(p+2)}$, while
if $p>2$ fluctuations take over and the velocity 
is controlled by the intrinsic roughness of a free
Edwards-Wilkinson, leading to a fluctuation-dominated regime
characterized by $\langle h \rangle \sim t ^{1/4}$
($0$, or logarithmic growth for two-dimensional interfaces).
These results are derived in a formal
way and extended in the appendix. \vspace{0.5cm}

\noindent{\it 2. Attractive potential: first order unbinding}.
For $b < 0$, by contrast to the complete wetting case, the surface
does not promote the growth of the liquid phase and consequently there
is no wetting phase even at bulk coexistence, $a=0$. $V(h)$ exhibits a
local minimum near the substrate, that binds the interface in the
presence of thermal fluctuations, and the width of the wetting layer
is finite (microscopic) at $a=0$.  We may, however, observe a
first-order unbinding transition that occurs as $a$ changes from
positive (stable bulk liquid) to negative (stable bulk gas) values.
\vspace{0.5cm}

\noindent{\it 3. Critical wetting}.
At a particular value of $b=b_c$ ($b_c=0$ in mean-field but more
generally $b_c$ is small and negative) critical wetting may be
observed, with a characteristic non-trivial phenomenology. This
situation requires the fine tuning of two independent parameters
($b=b_w$, $a=a_c$). This critical transition is more difficult to
treat theoretically and less likely to be found in real systems and
will not be discussed in this paper.
\vspace{0,5cm}

The best way to extend equilibrium approaches to more general,
non-equilibrium, situations is to consider the simplest and widely
studied non-equilibrium extension of the Edwards-Wilkinson equation,
i.e. the Kardar-Parisi-Zhang (KPZ) \cite{KPZ,HZ} interfacial dynamics
\cite{MBE}, in the presence of effective bounding potentials, as the ones we have 
described before.  This strategy has been followed in a series of
recent papers for systems with short-ranged (attractive and purely
repulsive) potentials \cite{reviews} and will be extended in the
present work to the case of long-ranged potentials.  We will discuss
the phase diagrams for both purely repulsive and attractive
potentials, paying special attention to criticality and to the
comparison with equilibrium wetting and non-equilibrium short-ranged
unbinding. We will focus on one-dimensional interfaces (separating
two-dimensional bulk phases), and mention briefly two-dimensional
interfaces in the conclusions.

The paper is organized as follows. In section 2 we introduce the
non-equilibrium unbinding model. In section 3, we review known results
for non-equilibrium short-ranged unbinding. Section 4 contains the
main body of the paper, including both analytical and numerical
results for purely repulsive and attractive potentials. Finally, the
main conclusions are presented together with a discussion of our
results.


\section{Non-equilibrium long-ranged unbinding: the model}

Our model consists in a KPZ non-equilibrium interface
\cite{KPZ,HZ} in the presence of a long-ranged, bounding potential
Eq. (\ref{LR}),
\begin{equation}
\partial_t h = \nabla^2 h +\lambda (\nabla h)^2 +a +
\frac{b}{h^{p+1}}+\frac{c}{h^{q+1}}+ \sigma \eta({\bf x},t),
\label{KPZwallLR}
\end{equation} 
%
where $\lambda \neq 0$ is the coefficient of the non-linear KPZ term,
the only new ingredient added to the equilibrium wetting Langevin
Eq. (\ref{EWwall}). 

Note that in equilibrium the time-dependent probability distribution
$P(h,t)$ is symmetric for the free interface and therefore it does not 
make any difference which side faces the substrate.  By contrast, 
under non-equilibrium conditions, owing to the
$h \to -h$ asymmetry of the KPZ-equation,
it depends on the sign of $\lambda$ 
that the substrate probes either one tail or the other of a 
KPZ probability distribution that is no longer symmetric.
Thus, for a given bounding potential
two different situations must be considered.
Therefore, we will investigate systems with positive and negative values 
of $\lambda$ (without loss of generality we take $\lambda= \pm 1$), 
and with both attractive ($b<0$) and repulsive
($b>0$) potentials, i.e. we consider {\it four distinct cases}. The
focus is mainly on one-dimensional interfaces.

For {\it analytical studies} we employ simple power-counting arguments
to establish the relevance or irrelevance of the new terms at the
equilibrium renormalization group fixed points. These will be combined
with heuristic and scaling arguments, to relate the emerging critical
behavior to equilibrium wetting and short-ranged non-equilibrium
unbinding.

For {\it numerical studies}, we consider one-dimensional
discretizations of Eq. (\ref{KPZwallLR}). As direct integrations of
KPZ-like equations are known to be plagued with numerical
instabilities \cite{newman-bray}, we resort to the exponential or
Cole-Hopf transformation, $n=\exp(\pm h)$, that leads to well-behaved,
numerically tractable, Langevin equations with multiplicative noise
\cite{reviews,MN}. In order to integrate these equations we employ a
recently proposed efficient numerical scheme \cite{DCM}, specifically
designed to deal with stochastic equations with non-additive
noise. More than just a useful technical trick, this transformation
has an interesting physical motivation, as we discuss next. For
negative values of $a-a_c$, the average interfacial height $\langle h
\rangle$ (thickness of the liquid film) may be large but finite, and
the interface fluctuates around its average position, occasionally
touching the substrate.  As the interface moves to infinity when $a
\to a_c$, its average height grows (i.e. the liquid film completely
wets the substrate) thereby suppressing contact (dry) sites. An
appropriate order-parameter (OP) for the unbinding transition is the number
of contact (dry) sites \cite{Haye-wetting,altenberg}, or equivalently
the surface order parameter \cite{surface}. This OP is
finite and positive when the interface is bound, and vanishes at the
unbinding transition. The variable, $ \langle n \rangle = \langle \exp(-h) \rangle$, 
that vanishes exponentially far from the wall, is an adequate mathematical
representation of such an OP (though not the only one).


The main goal of our study is the description of the scaling behavior
of the OP. $\langle n \rangle$, is expected to obey
simple scaling near the critical point, for sufficiently large times,
$t$, and large system-sizes $L$. Denoting $\delta a =|a-a_c|$,
\begin{equation}
\langle n (\delta a,t,L) \rangle= L^{-\beta_{OP}/\nu} \langle n  
(L^{1/\nu} \delta a,L^{-z} t) \rangle,
\label{scaling}
\end{equation}
while right at the transition $\langle n (\delta a=0,t) \rangle \sim
t^{-\beta_{OP}/\nu z}\sim t^{-\theta_{OP}}$ and therefore $\langle n
(\delta a,t=\infty) \rangle \sim \delta a^{\beta_{OP}}$, where the critical
exponents were introduced following standard nomenclature.
Analogously, for the interfacial height we can define $ \langle h
\rangle \sim \delta a^{-\beta_h}$ and $\langle h(\delta a=0,t) \rangle \sim
t^{\beta_h/\nu z}\sim t^{\theta_h}$, although
in terms of $h$ a single universality class,
with exponents related to the free KPZ \cite{MN}, is observed for both
signs of $\lambda$.
Determining all of these
critical exponents by the aforementioned techniques will
allow us to assign the emerging critical behavior to specific
universality classes, providing a comprehensive classification of
non-equilibrium unbinding transitions in the presence of long-ranged
forces. 

Before proceeding to the presentation of our results, we 
notice that it is expected that the behavior for short-ranged
interactions is recovered in the large-$p$ limit of
the long-ranged ones. Next, a brief review of the former 
is provided.

\section{Brief review of non-equilibrium short-ranged unbinding}

The KPZ equation with exponential bounding potentials is
\begin{equation}
\partial_t h = \nabla^2 h +\lambda (\nabla h)^2 + 
a +b~e^{-ph}+ c~ e^{-qh} + \sigma \eta,
\label{KPZwallSR}
\end{equation} 
with $q>p>0$. 
The results for the four possible physical situations are:

\subsubsection{Repulsive wall and $\lambda <0$}

\begin{figure}
\centerline{\psfig{figure=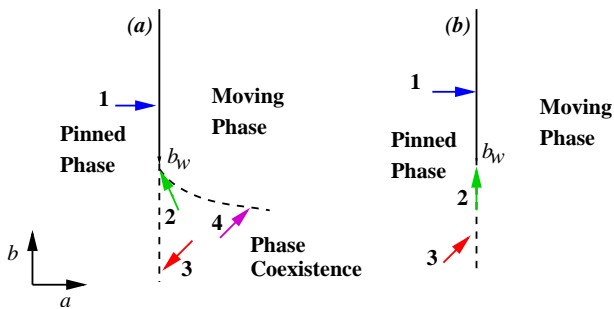,height=4.0cm}}
\caption{Phase diagrams for $\lambda <0$ $(a)$,
$\lambda >0$ ($b$). Paths labeled 1 correspond to
non-equilibrium complete wetting transitions; 2, critical or first-order
unbinding transitions (not studied); 3, first-order unbinding transitions;
4, unbinding transition in the directed percolation universality
class.  For $\lambda<0$ and $b<b_w$ (attractive substrates) ,
two-phase coexistence is observed in the area delimited by the two lines.}
\label{fig2}
\end{figure}

If $\lambda<0$ (we set $\lambda=-1$) the change of variables $n=\exp(-h)$
transforms (\ref{KPZwallSR}) (with $c=0$) into
\begin{equation}
\partial_t n = \nabla^2 n -a~n - b~ n^{1+p}  + n \sigma \eta.
\label{mn1}
\end{equation} 
This describes complete wetting transitions (along path 1 in Fig.
\ref{fig2}(a)) characterized by (see \cite{reviews}) a dynamic exponent
$z=3/2$, identical to KPZ, $\nu=1/(2z-2)=1$, and non-trivial exponents
$\beta_{OP}$ and $\theta_{OP}$ that were determined by
simulations. The exponents for $h$ have been measured also and the
transition was shown to be in the {\it multiplicative noise 1} (MN1)
universality class: $\beta_{OP} \approx 1.78$, $\theta_{OP} \approx
1.18$, $\theta_h \approx 0.33$ and $\beta_h \approx 1/2$ in $d=1$
\cite{reviews}.

\subsubsection{Repulsive wall and $\lambda >0$}

As for the $\lambda <0$ case, it is more convenient \cite{Omar} to use
the transformation, $n=\exp(+h)$ leading to
\begin{equation}
\partial_t n = \nabla^2 n + an + b n^{1-p} + n \sigma \eta.
\label{mn2}
\end{equation} 
This equation describes the transition along the path $1$ in Fig.
\ref{fig2}(b). Numerical estimates for the associated universality class have
been recently obtained from this (non-order-parameter) Langevin
equation \cite{Omar}. By measuring the order parameter $m=\langle 1/n
\rangle$ (that vanishes at the transition), the
following set of exponents was obtained: $\theta_{OP} \approx 0.22$,
$\beta_{OP} \approx 0.32$, different from MN1 and $ \theta_h \approx
0.33$, $\beta_h \approx 0.5$, $z=3/2$ and $\nu=1$ in line with the
corresponding exponents of the MN1 class
\cite{Omar,Fattah,Haye-wetting}. This universality class is known as
{\it multiplicative noise 2} (MN2).
A detailed discussion of the differences between the MN2 and MN1 universality 
classes, may be found in \cite{reviews}.

Note that, apart from the signs, the difference between
Eq. (\ref{mn1}) and Eq. (\ref{mn2}) is in the leading power of $n$. It
is possible, however, to summarize these two Langevin equations in
\begin{equation}
\partial_t n= \nabla^2 n + \alpha ~ a~ n + \alpha ~ b~ n^{\gamma} +n 
\sigma \eta,
\end{equation}
with $\alpha=\lambda/|\lambda|$ and $\gamma =1 - \alpha p$. Then $\alpha >1$ 
and $\alpha<1$ correspond, respectively, to the MN1 and MN2
universality classes. In the first case the leading power for large
values of $n$ is the non-linear term while this role is taken by 
the linear term in the second case. The transition at the boundary
$\gamma = 1$ ($p=0$) is obviously discontinuous, as both terms
are linear and there is no saturating term.

In MN1 the order parameter is $n$, while in the MN2 case, it is $m=1/n$. 
In both cases $a$ is the control parameter.

\subsubsection{Attractive wall and $\lambda <0$}

For attractive walls, $b < 0$, a positive value of $c$ is required for
stability, for any value of $\lambda$. In systems with $\lambda <0$
(see Fig. 1(a)), a new phenomenology including a broad coexistence
region, and a directed-percolation unbinding transition emerges
\cite{synchro,reviews,broad,Haye-wetting}. In the broad-coexistence
region the stationary solution is either bound or unbound depending on
the initial conditions \cite{reviews,Lisboa}. Such a region is
delimited on the right (where the bound phase loses stability) by a
directed percolation transition, where the scaling properties are
controlled by the effective dynamics of the particle-like
interface-surface contact points (i.e. points trapped in the potential
well). Its leftmost border corresponds to the abrupt (discontinuous)
binding of initially unbound interfaces. Again we refer the reader to
\cite{reviews} for a detailed discussion and to \cite{broad} for
a review on generic phase-coexistence in non-equilibrium systems.

\subsubsection{Attractive wall and $\lambda >0$}

For $\lambda >0$ (see Fig. 1(b)) a first-order transition separates
bound from unbound phases (akin to the equilibrium discontinuous
transition for attractive walls). No broad coexistence region, nor directed
percolation transition, exist in this case.

\section{Non-equilibrium long-ranged unbinding: results}

\begin{figure}
\centerline{\psfig{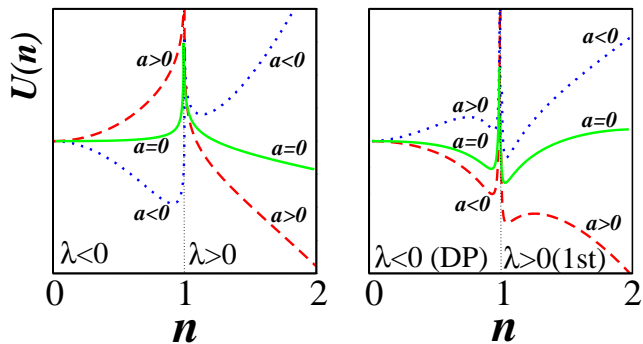}}
\caption{Effective potentials in terms of $n$ obtained from the 
numerical integration of the interaction part of eq. (\ref{MNLR}).
Left and right panels correspond to, respectively, repulsive ($b=1$) and
attractive ($b=-1$) interactions.  
For both $b$ we show results for negative and positive $\lambda$, and 
different values of $a$ corresponding to the bound and unbound phases, 
as well as at coexistence. 
Note that for $\lambda<0 (>0)$ the unbound phase corresponds to 
a minimum of $n$ at 0 ($\infty$). Transitions for attractive walls 
can be either first-order or continuous (directed-percolation).} 
\label{fig3}
\end{figure}

We are now set to discuss the long-ranged non-equilibrium problem
described by Eq. (\ref{KPZwallLR}).
%
%
There is a singularity at $h=0$ and thus only positive values of
$h$ are allowed (mimicking the impenetrability of the substrate). As before, 
if $b>0$ we take $c=0$ for simplicity. Proceeding as in the
short-ranged non-equilibrium case, we perform the change of variables
$n=\exp(\alpha h)$, with $\alpha = \lambda/|\lambda|$, in equation
(\ref{KPZwallLR}), obtaining
\begin{equation}
\partial_t n = \nabla^2 n +\alpha a n +\alpha b
\frac{n}{|\alpha \log(n)|^{1+p}} + n \sigma \eta,
\label{MNLR}
\end{equation} 
where a term $+\alpha c n/|\alpha \log(n)|^{1+q}$ has to be added when
$b<0$. As before, for positive $\lambda$ ($\alpha=1$), the
order-parameter is $m=1/n$, while for $\lambda <0$ the order-parameter
is $n$ itself. Note also that as there is a singularity at $n=1$
(inherited from the singularity at $h=0$ in Eq. (\ref{KPZwallLR})),
for $\lambda>0$, where $n$ diverges at the transition, the initial
condition is fixed at $ n(x) > 1 ~~ \forall x$, while for $\lambda
<0$, where $n$ vanishes at the transition, $0 < n(x) < 1 ~~\forall x$,
is taken. The deterministic one-site terms of Eq. (\ref{MNLR}) may be
written as minus the derivative of a potential, $U(n)$, that is
depicted in Fig. \ref{fig3}.

It is instructive to compare this model with the two universality
classes reported for non-equilibrium short-ranged wetting, i.e. MN1
and MN2. In fact, it is expected that, in the limit of sufficiently
large $p$, the power-law force yields the same dynamics as
short-ranged (exponential) forces. Thus, for $\lambda<0$ and large $p$
we anticipate MN1 behavior while MN2 scaling should obtain when
$\lambda > 0$, in the same limit.

\subsection{Analytic results}

In an early work the KPZ non-linearity was argued to be irrelevant
above the (mean-field) wetting temperature $b=b_w=0$, and an
equilibrium (complete) wetting transition was predicted to occur as $a
\to a_c$, at constant $b>b_w$, for any $\lambda$ \cite{Lipowsky_kpz}
(transitions along path 1 in Fig. \ref{fig2}). In the following
we show that such a prediction is untenable and that the
non-equilibrium term leads to new physics.

Let us start by employing na{\"\i}ve power counting arguments, based
on equilibrium scaling, to decide whether $\lambda$ is a relevant or
an irrelevant perturbation, at the mean-field fixed point and at the
fluctuation one. In order to do that, we first fix $\lambda=0$ in
equation (\ref{KPZwallLR}). If $b>0$, then the upper critical
dimension depends only on the repulsive part of the potential and is
$d_c(p)=2p/(2+p)$ \cite{Lipowsky84,GF}, as shown in the appendix.
Now, from dimensional analysis $[\lambda] = L^{-1+d/2}$. Upon
evaluating it at $d_c(p)$ one finds $[\lambda] =L^{-2/(2+p)}$ which,
in terms of momenta, has a positive dimension for any value of
$p$. Therefore, the KPZ non-linearity is relevant at the mean-field
equilibrium wetting transition.

Relevancy at the fluctuating regime fixed-point is proved using the
known one-dimensional scaling dimension of the field $[h] \sim t^{1/4}
\sim L^{1/2}$ at the fluctuation-dominated fixed point (see appendix). 
Then, it
follows, $[\lambda] =L^{-(d+1)}$, implying that $\lambda$ is strongly
relevant in any space-dimension. To be rigorous we would need to include
perturbative corrections generated by the new non-linear term
proportional to $\lambda$, but even without computing these, one can
say that it is very unlikely that such corrections reverse the strong
lowest-order relevancy of $\lambda$.  The relevancy of $\lambda$, is
strongly supported by the results of numerical simulations of the
corresponding Langevin equation as we will show next.

As in one-dimensional equilibrium interfaces, where $p=2$ separates
the mean-field and the fluctuation-dominated regimes, it is easy to
argue that in non-equilibrium the two regimes are separated by $p=1$.
From Eq. (\ref{KPZwallLR}) in the absence of noise, the mean-field
velocity exponent at the critical point, given by $\lambda
\langle (\nabla h)^2 \rangle + a_c=0$, is obtained by integrating
$\partial_t h \sim h^{-p-1}$, and found to be $\theta_h=1/(p+2)$. On
the other hand, when noise (fluctuations) is included, the
(one-dimensional) free KPZ equation has a roughening exponent of $1/3$
and, therefore, a velocity proportional to $t^{1/3}$
\cite{HZ}. Which of these contributions dominates? Clearly, if
$p<1$ the wall-induced velocity is larger and fluctuations give only a
higher order correction (i.e. they are irrelevant). By contrast, if
$p\geq 1$ the effective repulsion generated by the wall (through
suppression of the intrinsic interfacial roughness) controls the
scaling. Thus, in non-equilibrium long-ranged wetting, $p=1$ separates
the mean-field from the fluctuation-dominated regimes.

Transient effects, that are significant before the non-equilibrium
interface develops its full (asymptotic) time-dependent roughness, may
prevent the KPZ exponent $\theta=1/3$ from being observed, leading to
an effective exponent, $\theta_{eff}<1/3$. Furthermore, at short
times, the interface is expected to grow with an Edwards-Wilkinson
exponent, $\theta=1/4$, and therefore $\theta_{eff}$, increases
progressively from $1/4$ to its asymptotic KPZ value, $1/3$, in the
long time regime. Comparing these values with the wall induced
velocity exponent $1/(p+2)$, we anticipate that for potentials with
$1< p < 2$ severe transient effects will occur before the
fluctuation-dominated scaling sets in. By contrast, for $p> 2$
fluctuations dominate from the early stages of interfacial growth.

\subsection{Numerical Results}

In order to avoid numerical instabilities, typical of KPZ direct
numerical integration schemes \cite{newman-bray}, we chose to study
the associated multiplicative noise Eq. (\ref{MNLR}) obtained after
performing a Cole-Hopf transformation. To solve Eq. (\ref{MNLR})
efficiently we have used a recently proposed {\it split-step} scheme
for the integration of Langevin equations with non-additive noise
\cite{DCM}. In this scheme, the equation under consideration is
discretized in space and time and separated into two contributions:
(i) the first includes deterministic terms only and is integrated at
each time-step using a standard integration scheme: Euler, Runge
Kutta, etc \cite{Maxi} (here we have chosen a simple Euler algorithm)
(ii) the output of the first step is used as input to integrate (along
the same discrete time-step) the second part which includes the noise
and, optionally, linear deterministic terms. This is done by sampling
the probability-distribution, i.e. the solution of the Fokker-Planck
equation associated with this part of the dynamics. In the case under
study (noise proportional to the field), the second step can be
carried out exactly. At each site, one has to sample a log-normal
distribution, i.e., the solution of the Fokker-Planck equation
associated with $\partial_t n= \alpha a n + \sigma n \eta$ (for more
details see \cite{lognorm} and \cite{DCM}). The two-step algorithm for
Eq. (\ref{MNLR}) is then implemented as follows.  At each site
$n=n(x,t)$, we compute
\begin{eqnarray}
n_1(x,t)& = & n+dt \left[ \frac{\alpha b~n}{\big(\alpha \log(n)\big)^{1+p}}
+\nabla_{discr}^2 n(x,t) \right]
\end{eqnarray}
where the discretized Laplacian is defined by

\begin{equation}
\nabla_{discr}^{2}
n(x,t)= \frac{ n(x+\Delta x,t)+n(x-\Delta x,t)-2n(x,t)}{\Delta x^2}
\end{equation}
with $\Delta x$ the space-mesh, and
\begin{equation}
n(x,t+ \Delta t)=n_1(x,t)~ \exp\left(\alpha a \Delta t + \sigma~ \eta~
\sqrt{\Delta t} \right)
\end{equation}
where $\eta$ is a random variable extracted from a Normal distribution
with zero-mean and unit variance. Note that the linear deterministic
term can be included in either the first or the second step, or
partially incorporated in both of them. For systems with $b<0$, the
stabilizing term, proportional to $c$ has be to included.

We set $\sigma=1$, $\Delta x=1/\sqrt{0.1}$, and the time-mesh $\Delta t
=0.1$ (note that in this scheme $\Delta t$ can be taken larger than in
the usual integration algorithms \cite{DCM}). In some simulations we used 
different values of $b$, which by default was set to $b=\pm 1$. We take 
as initial condition $n(x,t=0)= 3$ if $\lambda >0$ (recall that $n \in ]1,\infty[$) 
and $n=0.5$ if $\lambda <0$ ($n \in [0,1[$). Then, the dynamics is iterated 
by employing the two-step integration algorithm at each site and using parallel
updating.

The numerical procedure is as follows. In order to determine the
critical point for any set of parameters we take the system-size as
large as possible and look for the separatrix between upward-bending
and downward-bending curves in the order-parameter (either $n$ or
$m=1/n$ depending on the case) versus $t$ in a double-logarithmic
plot. The asymptotic value of this slope gives an estimation of
$\theta_{OP}$. Also, for the same parameters, $\langle h \rangle$
grows as a power-law with an exponent $\theta_h$ (bending downward and
upward in the bound and the unbound phases, respectively). Generally
the order parameter is more sensitive to control-parameter variations,
providing the most reliable way of determining the critical point. For
completeness, and in order to check the validity of analytical
approximations, we measure the global interface width, $W$, at the
transition, which is expected to grow with the KPZ exponent
$\beta_W=1/3$, in the regime where it is asymptotically free.

Once the critical point is determined accurately we compute
$\beta_{OP}$ and $\beta_h$ by measuring the stationary values of the
order parameter and of $\langle h \rangle$ at different distances from
it. A complementary approach is based on finite-size scaling
analysis: the values of the order parameter and of $\langle h
\rangle$, at saturation, are measured for a fixed value of $a$ 
as a function of system-size. At the critical point these values scale
with exponents $\beta_{OP}/\nu$ and $\beta_h/\nu$, respectively. In
addition, the scaling of the saturation times for different
system-sizes allows to determine the dynamical exponent $z$.  This
standard finite-size scaling analysis is not always possible (see
below), and in such cases $z$ is measured through spreading-like
experiments. Finally, an alternative to spreading consists in
measuring the distribution of gaps between contact points at a given
time. For small gaps this function decays with an exponent
$z\theta_{OP}$ giving yet another estimate of $z$ \cite{FSS-MN1}.
 
The correlation length critical exponent $\nu$ is obtained by
measuring the location of the effective critical point, i.e. the value
of $a$ for which the order-parameter falls below a fixed threshold,
say $10^{-3}$, as a function of system-size: $a_{c,eff}(L) \sim
L^{-1/\nu}$. This exponent may also be determined indirectly by
employing scaling relations and using the value of $\beta_{OP}/\nu$
from finite-size scaling analysis and the value of $\beta_{OP}$
obtained from direct measurements.  The results of these measurements,
in conjunction with the scaling laws, provide an over-complete
estimation of the set of critical exponents, that was also used to
verify scaling relations.

Before discussing the differences between the various universality classes 
and regimes (i.e. different values of $\lambda$, $p$ and $b$) we first give 
an overview of the common features of all simulations.

\begin{enumerate}

\item Once the KPZ equation parameters
($D, \lambda, \sigma$) are fixed, the location of the critical point is
universal, meaning that it does not depend on the details of the substrate, i.e. 
on the values of $b, c$, and $p$. The critical point is determined by the
value of $a$ where the free-KPZ interface changes the sign of its velocity, 
from positive, i.e. diverging to an unbound state, to negative,
becoming bound at the wall: $a_c + \lambda \langle (\nabla h)^2
\rangle =0$. As we consider two different values of $\lambda$, $+1$
and $-1$, there are two critical points: 
$a_c(\lambda = \pm 1) \approx \mp 0.143668(3)$. 

\begin{figure}
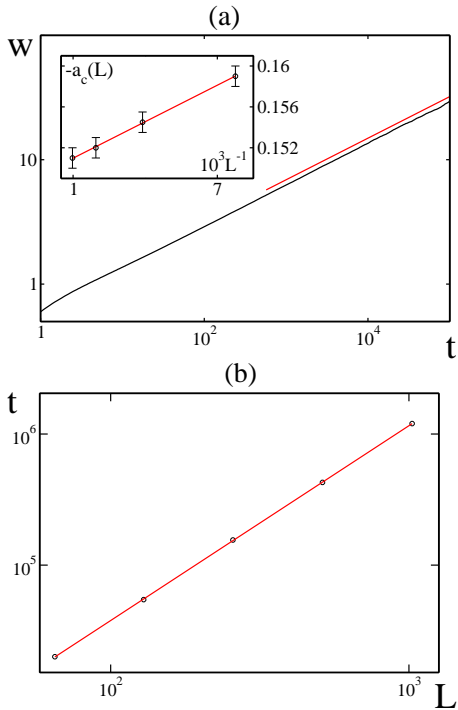

\begin{center}
\includegraphics[angle=0,width=6cm]{fig4a.eps}
\includegraphics[angle=0,width=6cm]{fig4b.eps}
\caption{Features common to all simulations. (a)
Roughness vs $t$ gives $\beta_W=0.33(1)$; (inset) $-a_c(L)$ vs
$10^3L^{-1}$ falls on a straight-line that yields $\nu \sim 1$ (data for
$\lambda=-1$ and $p=2$). (b) Saturation time vs system-size leads to
$z=1.48(4)$ (data for $\lambda=1$ and $p=2$).}
\label{fig4}
\end{center}
\end{figure}

\item At the critical point, the asymptotically unbound interface is a 
free KPZ one, and thus $z=3/2$ and $\beta_W = 1/3$. These values were 
consistently checked in all simulations (see Fig. \ref{fig4}(a),(b)).

\item A simple argument, originally given in \cite{MN}, predicts $\nu=1$ 
for all bounded KPZ interfaces. This prediction was confirmed in all of 
our cases (see inset of Fig. \ref{fig4}(a)).

\end{enumerate}

\subsubsection{Repulsive walls and $\lambda > 0 $.}

We have to distinguish two regimes, depending on the
range of the attractive substrate, i.e. the value of $p$.
\vspace{0.5cm}

\begin{figure}
\begin{center}
\includegraphics[angle=0,width=6cm]{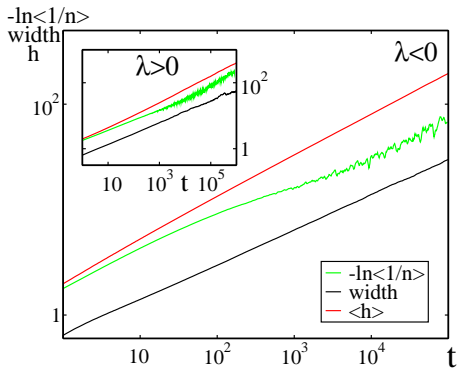}
\caption{Log-log plot of the time evolution at $a_c$ of $\langle h
\rangle$ (upper, red curves), $-\ln \langle n_{OP}\rangle$ (middle,
green curves), and the width $w$ (lower, black curves), in the
mean-field regime $p=0.5$, for $\lambda<0$ (main) and $\lambda>0$
(inset). Irrespective of the sign of $\lambda$, $\langle h \rangle$
and the roughness may be fitted to a power-law with the predicted
exponents $\theta_h = 1/(p+2)$ and $\beta_W=1/3$, respectively.  $-\ln
\langle n_{OP}\rangle$ falls on a straight-line in a double
logarithmic plot, confirming the stretched-exponential behavior of the
order parameter.  }
\label{fig5}
\end{center}
\end{figure}

\noindent{\it Mean-field regime.}
The theoretical discussion indicates that for $p<1$, and any sign of
$\lambda$, a mean-field regime controlled by the exponents
$\theta_h=1/(p+2)$ and $\beta_h=3/(2p+4)$ is obtained. By changing
variables in a na\"ive way a stretched exponential behavior for the
order-parameter is predicted. Figures
\ref{fig5} and \ref{fig10} illustrate the confirmation of these
predictions (both for positive and negative $\lambda$).

\begin{figure}
\begin{center}
\includegraphics[angle=0,width=8cm]{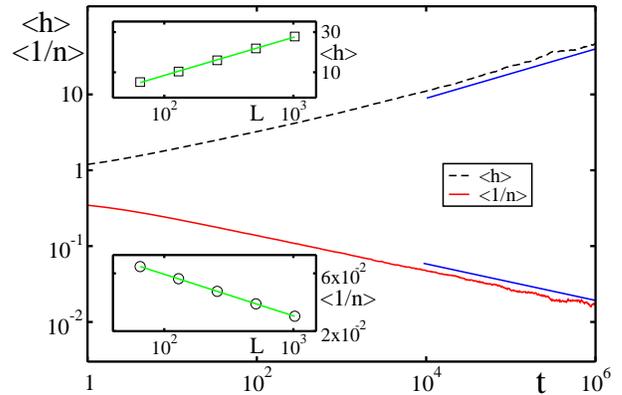}
\caption{Log-log plot of the time evolution at $a_c$ of $\langle h \rangle$
(dashed, black curve), and $-\log \langle n_{OP}\rangle$ (solid, red
curve), in the fluctuation regime ($p=2$) and for $\lambda>0$. From
the slopes of the straight-line fits one finds $\theta_{h}
=0.32(2)$(upper-curve) and $\theta_{OP} = 0.228(6)$
(lower-curve). Upper-inset: finite-size scaling of $\langle h \rangle$
yielding $\beta_{h}/\nu = 0.46(2)$. From the lower-inset one obtains
$\beta_{OP}/\nu = 0.34(2)$. These exponents agree with those of the
MN2 universality class.}
\label{fig6}
\end{center}
\end{figure}

\noindent{\it Fluctuation regime: Multiplicative Noise 2.}
A strong-fluctuation regime is predicted for systems with $p>1$ but,
as argued above, severe transient effects are expected for $2>p>1$. We
start with the analysis of the, a priori, simpler $p>2$ sub-regime and
offer simulation results for $p=2, 2.5, 3, 4, 7$. In all cases the
order-parameter was found to decay at criticality with an exponent
$\theta_{OP} \approx 0.229 $ while the average height diverges with
$\theta_h \approx {1/3}$ (see Fig. \ref{fig6}, data shown for
$p=2$). A standard finite-size scaling analysis can be performed (see
Fig. \ref{fig6}), yielding $\beta_{OP}/\nu = 0.34(2)$ and
$\beta_h/\nu = 0.46(2)$.

These results, together with the previously reported general ones,
unambiguously place the fluctuation regime for repulsive
walls with positive $\lambda$ into the MN2 universality class.

\begin{figure}
\begin{center}
\includegraphics[angle=0,width=8cm]{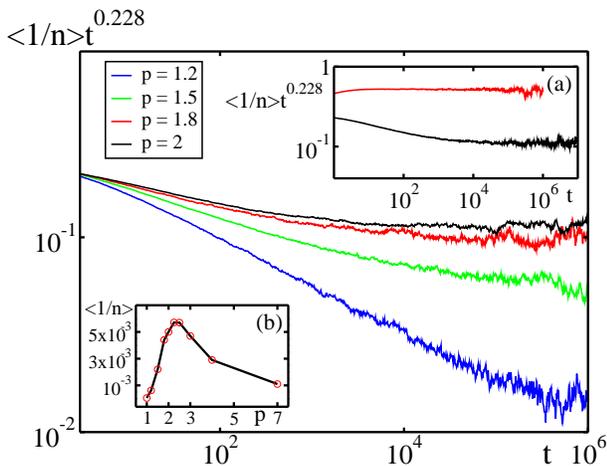}
\caption{Main: Order-parameter multiplied by the expected power-law
$t^{0.228}$. For systems with $b=1$, long transients that depend on
$p$ are observed for systems with $1<p<2$. Inset (a): The crossover
times are reduced as $b$ decreases; compare the upper, red curve for
$b=0.05$ with the lower, black one for $b=1$ ($p=2$). Inset (b):
Order-parameter at $t=10^6$ vs $p$ ($b=1$). $p=2$ marks the boundary
of the strong transient region as illustrated by the different
behaviors observed above and below $p=2$.}
\label{fig7}
\end{center}
\end{figure}

For systems with $1<p<2$, where strong transients are expected, after
fixing $b=1$ and running simulations up to $t=10^6$, continuously
varying power-law exponents are found (see Fig. \ref{fig7}). We
note, however, that these fits give effective rather than asymptotic
exponents. In fact, the change in the effective exponents from
mean-field (wall-controlled) to the fluctuation (intrinsic-interface)
regime is expected to occur at shorter times when the effect of the
substrate is less pronounced, implying that reducing the value of $b$
decreases the crossover time. This was confirmed by simulating systems
with $b=0.1$ and $b=0.05$ and observing a monotonic decrease of the
effective exponents that converge to the expected asymptotic value
$\theta_{OP} \approx 0.228$, $\theta_{h} \approx 1/3$ (see inset (a)
of Fig. \ref{fig7}) in line with the hypothesis that the
transition belongs to the MN2 universality class.

In order to check that $p=2$ is the boundary between the strong and
weak transient sub-regimes, we have plotted in Fig. \ref{fig7},
inset (b), the average order-parameter for systems with the same
initial condition, at time $t=10^6$, and different values of $p$. This
is a non-stationary value of the OP that is strongly affected by
transients. It is clear from the figure that the behavior of the order
parameter changes qualitatively at $p=2$ corroborating the result that
this value of $p$ marks the boundary between the sub-regimes with and
without severe transients.

\subsubsection{Repulsive walls and $\lambda < 0 $.}

\noindent{\it Mean-field regime.}
In parallel with the positive $\lambda$ case, the results of Figs.
\ref{fig5} and \ref{fig10} show that the theoretically predicted
mean-field regime is clearly observed for systems with $p<1$.
\vspace{0.5cm}

\begin{figure}
\begin{center}
\includegraphics[angle=0,width=8cm]{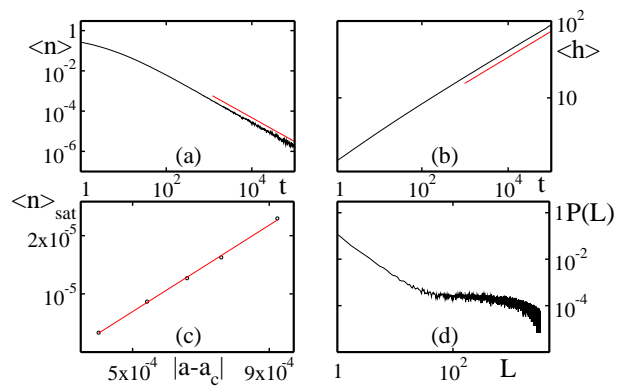}
\caption{Results for $\lambda<0$ and $p=2$. 
(a) Log-log plot of the
time decay of the order-parameter. The straight-line is a guide to the
eye, and has a slope $\theta_{OP}=1.19(1)$.
(b) Log-log plot of the average
distance from the wall vs time, leading to $\theta_{h}=0.33(1)$.
(c) The scaling of the saturation value of the order-parameter yields
$\beta_{OP}=1.76(3)$.
(d) Log-log plot of the gap (between contact points) distribution function. 
The initial slope gives $z\theta_{OP} \approx 1.75(10)$.}
\label{fig8}
\end{center}
\end{figure}

\noindent{\it Fluctuation regime: Multiplicative Noise 1.}
Again we have to distinguish two sub-regimes, with and without 
severe transients, depending on whether $p$ is larger than or smaller
than $2$. Simulations in the weak transient regime were performed for $p=2$ and $3$. 
In both of these systems the order-parameter 
decays at criticality with an exponent $\theta_{OP} = 1.19(1) $ while the
average height diverges with $\theta_h = 0.33(1)$ (see Fig.
\ref{fig8}(a),(b) data shown for $p=2$). 
As was first pointed out in \cite{FSS-MN1}, finite-size scaling
measurements are non trivial in this case due to the presence of two
different characteristic times. Namely, the correlation length reaches
the size of the system at times $\sim L^z$, whilst the interface
typically detaches from the wall at times $\sim L^{1/\theta_n}$.  As
the latter grows with a larger exponent for MN1, the interface
detaches from the wall before it reaches the saturation regime for
finite samples, rendering the evaluation of $\beta_{OP}/\nu$ and $z$
through standard finite-size scaling methods problematic. An
estimation of $\beta_{OP}$ is possible by taking a large system-size,
$L=2^{17}$, and measuring the order-parameter stationary-state value
upon approaching $a_c$. We find $\beta_{OP} = 1.76(3)$ and $\beta_h =
0.51(3)$(see Fig. \ref{fig8}(c)). $z$ is accessible through
spreading experiments from an initial condition with only one active
(pinned) site. The measurement of the mean-square deviation from the
origin $R^2(t) \approx t^{2/z}$, gives $z=1.52(5)$ (not
shown). Alternatively, one can investigate the gap distribution
function of the distances between neighboring contact points at a
given time or, equivalently, the average size of inactive islands in
the $n$-language \cite{FSS-MN1}. For small gaps this function decays
with an exponent $z\theta_{OP}$, and we find $z\theta_{OP}=1.75(10)$
which leads to a value of $z$ compatible with 3/2(see Fig. \ref{fig8}(d)).

These results, together with the general ones, place unambiguously
this fluctuation regime for repulsive walls with negative
$\lambda$ into the MN1 universality class.

Again, for systems with $1<p<2$, different effective exponents are obtained 
at a fixed maximum time for different values of $b$ (unity and smaller), 
confirming the existence of strong transients.
Upon decreasing $b$, the influence of the wall is 
reduced and a behavior compatible with the MN1 class is observed: 
$\theta_{OP} \approx 1.19 $, $\theta_h \approx 1/3$, $\beta_{OP} \approx 1.76$ 
and $\beta_h \approx 0.5$ (figure not shown).

\subsubsection{Attractive wall and $\lambda < 0 $.}

The phase-diagram, depicted on the left panel of Fig.
\ref{fig2}, is similar to that found for short-ranged
interactions \cite{reviews}. For a fixed $b$, by varying $a$ one of
two transitions may occur depending upon the initial interfacial
state.  Initially unbound interfaces experience an unbinding-binding
transition at $a_c$ where the free-interface velocity inverts its sign
(in full analogy with the previous case; see path 3 in Fig.
\ref{fig2}). 
On the other hand, initially bound interfaces unbind at a different
non-trivial value of $a$, noted $a^* > a_c$, inside the free-interface
unbound phase (path 4 in Fig. \ref{fig2}(a)).  This transition is
analogous to the one observed for short-ranged forces, and is expected
to be controlled by the unbinding of interface-sites trapped in the
potential minimum. Bound sites (located around the potential minimum)
are identified with particles; unbound sites are described by holes.
The effective particle dynamics is very similar to that of the
contact-process
\cite{review_AS} (a well-studied model known to be in the directed
percolation class): an occupied site can become empty when a point
is detached, and can induce also the binding of a neighboring
site. Furthermore, empty sites cannot become spontaneously occupied 
in the absence of occupied (bound) neighboring sites. Indeed, as soon as
the interface is locally out of the potential well, it is pulled away
from it. This corresponds to the absorbing state characteristic of
the directed percolation class. Note that the statistics of the
average number of such pseudo-particles is completely analogous to 
that of $\langle \exp(-h) \rangle$.
 
Before the depinning transition, typical {\it triangular structures}
are observed, consisting of pinned sites (lying in the potential
well), and depinned sites being pulled from the substrate. This
triangular shapes (pyramidal in two-dimensions) are similar to those
in the analogous short-range case, and are reminiscent of pyramidal
mounds obtained in the non-equilibrium growing of some interfaces, as
for instance, in the so called Stranski-Krastanov effect \cite{SK}.

\begin{figure}
\begin{center}
\includegraphics[angle=0,width=8cm]{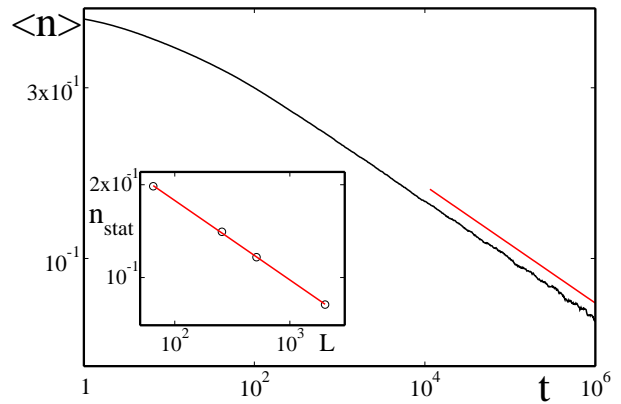}
\caption{Time decay of the order-parameter at the critical point
$a=0.38448$.  From the slope in the log-log plot we find
$\theta=0.161(2)$. Inset: finite size scaling of the order
parameter. From the slope in the log-log plot we estimate
$\beta/\nu=0.26(2)$, in agreement with directed percolation
values.}
\label{fig9}
\end{center}
\end{figure}

Our numerical results show that this transition is
controlled, as in the short-ranged case, by directed percolation 
critical exponents (see Fig. \ref{fig9}). In particular, we have
determined $\beta_{OP}/\nu=0.26(2)$ and $\theta_{OP}=0.161(2)$, in excellent agreement
with the one-dimensional directed percolation values.

Let us remark that, as the bound sites remain inside the potential
well, and the dynamics controlling their final ``escape'' is likely to
be insensitive to the exponential or power-law tails of the potential
at large values of $h$, the parallel between this behavior and the
directed-percolation transition for short-ranged forces is to be
expected.  Interestingly, Ginelli et al. investigated a lattice model
of a generalized contact process with long-ranged interactions between
the edges of low-density segments and found a transition in the
directed percolation universality class for forces that decay
sufficiently slowly, and a first-order transition otherwise
\cite{ginelli}. Clearly, in terms of $h$ this translates into a
long-ranged interaction between the vertices forming the triangle
bases, and it is reasonable to assume that, in turn, an effective
long-ranged attraction between the substrate and the interface must be
obtained. In the light of these results it is reasonable to assume,
that both short- and long-ranged interactions in similar models will
be characterized by the same behavior below $b_w$.

In the region between $a_c $ and $a^*$ one observes generic (broad)
phase coexistence: the stationary solution is either bound or unbound
depending on the initial condition. Within this region, the bound
phase is characterized by some bound sites trapped in the potential
minimum, and pseudo-unbound regions separating them \cite{reviews}.
In full analogy with short-ranged forces, close to the unbinding
transition $a \lesssim a^*$ initially bound interfaces are stable
owing to a mechanism that eliminates local fluctuations into the
unbound phase: once formed, islands of the unbound phase rapidly
transform into triangular mounds of fixed slope, which subsequently
shrink from the edges.

\subsubsection{Attractive wall and $\lambda > 0 $.}

When $\lambda > 0$, the situation is rather similar to the one
for equilibrium and for non-equilibrium ($\lambda > 0$) short-ranged
systems. At the critical value $a_c$ where the free interface inverts its
velocity sign, there is a discontinuous unbinding-binding
transition (path $3$ in Fig. \ref{fig2}). This value does not
depend on the value of $p$ nor on $b$ or $c$ \cite{transient}.

\vspace{0.5cm}

\noindent{\it The multicritical point}.
Finally, for either sign of $\lambda$, path 2 in Fig.
\ref{fig2} corresponds to a multi-critical point analog
to an equilibrium critical wetting transition when the critical 
point is approached at coexistence. 
Most likely, its location will not coincide with its
mean-field value $b=0$, but exhibits some renormalization shift. The
analysis of this multi-critical point will be considered elsewhere.

\subsection{Discrete Model}

As a final check of universality issues, we simulated a
discrete interfacial model, known to belong to the KPZ class, in the
presence of a long-ranged substrate. The model is the same as that
studied in the context of short-ranged wetting in \cite{Fattah}. Even
if plagued with long transient effects (much larger than in the
short-ranged case) all of the previously reported phase diagrams and
universality classes seem to be confirmed for
the different types of walls (i.e. values of $b$ and $p$) and signs of
the non-linearity. Generally, the discrete model provides slightly better 
results for the height variable as compared with the 
continuum model, and worse for the order-parameter.

\begin{figure}
\begin{center}
\includegraphics[angle=0,width=8cm]{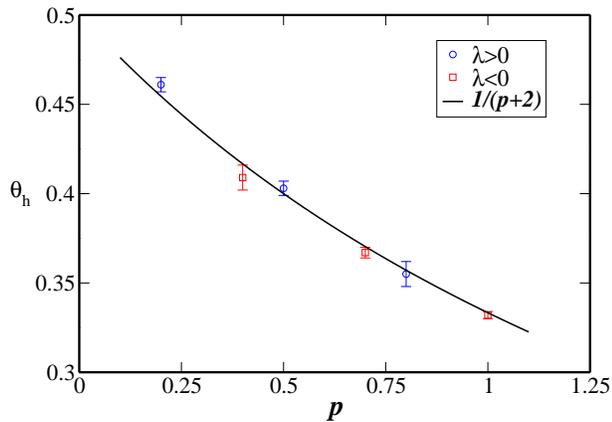}
\caption{Time growth exponents $\theta_h$ in the mean-field like
regime for $\langle h \rangle $ at the critical point,
as results from the discrete interfacial model. 
Blue circles(red squares) stand for $\lambda>0(<0)$ data
points, and the solid line is the predicted curve $1/(p+2)$.}
\label{fig10}
\end{center}
\end{figure}

Figure \ref{fig10} displays the time growth of the 
mean separation $\langle h \rangle$ in the mean-field like
regime ($p<1$), for both positive and negative $\lambda$.
Additionally, the ratio $\beta_{OP}/\nu=0.251(2)$ and $\theta_{OP}=0.156(2)$ were 
obtained for the directed percolation transition, which compares favorably
with the accepted estimates 0.25208(5) and 0.1595 \cite{jensen}.

\begin{figure*}
\begin{center}
\includegraphics[angle=0,width=16cm]{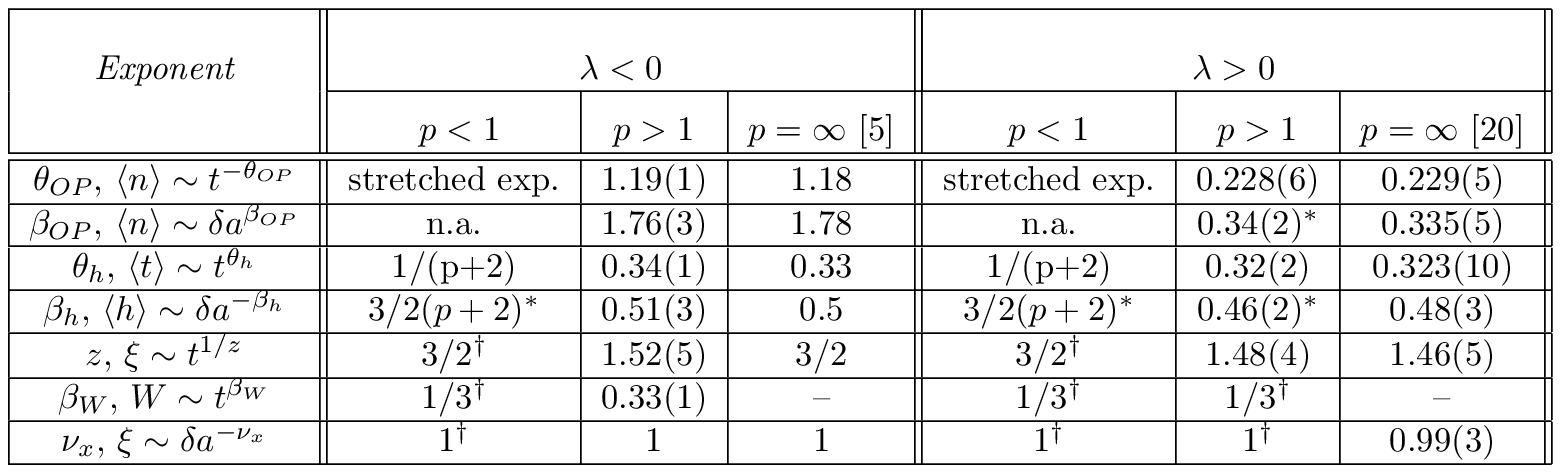}
\caption{Summary of the critical exponents in the 
mean-field ($p<1)$ and the fluctuation ($p>1$) regimes for non-equilibrium, 
complete wetting transitions with long-ranged forces. 
To facilitate the comparison, the exponents for the 
MN1 and MN2 universality classes are also included ($p=\infty$).
$^*$, exponent from finite-size analysis or scaling relations; 
$^\dagger$, estimated value from short simulations; 
n.a., not applicable.}
\label{table}
\end{center}
\end{figure*}

\section{Discussion and Conclusions} 

We have studied the unbinding of KPZ interfaces in the presence of
limiting substrates, interacting via long-ranged potentials. This is
the simplest model for interfacial effective descriptions of
wetting and in general, unbinding transitions, of systems
interacting through van der Waals forces under non-equilibrium
conditions.

We have presented the results of systematic analytical and numerical 
studies of one-dimensional KPZ-like interfaces in the presence of 
long-ranged forces Eq. (\ref{KPZwallLR}), supporting the following 
conclusions:

{\bf i)} Repulsive interactions drive a non-equilibrium complete
wetting transition for either sign of $\lambda$.  This transition
belongs to different universality classes depending on the strength of
the repulsion, i.e. on the value of $p$ in Eq. (\ref{KPZwallLR}) and
on the sign of $\lambda$. For $p<1$ a mean-field like regime is
observed in both cases, while for $p>1$ the fluctuation regime obtains and the 
transition is in the multiplicative noise 1 (MN1) class for $\lambda <0$ and in the
multiplicative noise 2 (MN2) for $\lambda > 0$. Systems in the
fluctuation regime, exhibit severe crossover effects for bounding
potentials with $1<p<2$. This should be contrasted with the behavior
of equilibrium systems where the value of $p$ that separates the mean
field from the fluctuation regimes was found to be $p=2$. More
importantly, in non-equilibrium systems the symmetry of the
equilibrium wetting and drying transitions is broken and the
fluctuation regime of the corresponding equilibrium wetting
transitions is split into two different non-equilibrium universality
classes, MN1 and MN2 respectively. Our results are collected in table
\ref{table}.

{\bf ii)} For attractive walls, i.e. below the critical wetting
temperature, phase-diagrams analogous to those of systems with 
short-ranged forces have been found: generic phase-coexistence over a 
finite area limited by directed percolation and first-order boundary 
lines for $\lambda <0$, and a first-order phase transition from an unbound
to a bound interface for $\lambda > 0$. This transition should 
not be called ``wetting'' as the interface detaches below the wetting 
transition temperature.

The unbinding transition at the critical wetting point (which in the
language of this paper corresponds to a multicritical point) requires
a higher degree of fine-tuning and is therefore expected to be more
difficult to observe in experimental situations. Its study is also more
laborious and is deferred to future work.

For more realistic two-dimensional interfaces, corresponding to
three-dimensional bulk systems, the situation is expected to be very
similar: all universality classes (mean-field, multiplicative noise 1,
multiplicative noise 2, and directed percolation) are expected to be
substituted by their two-dimensional counterparts, with analogous
phase diagrams and overall phenomenology.

We hope that the results described in this paper will help to
motivate an experimental study of wetting and unbinding transitions 
under non-equilibrium situations. In these systems one expects to find 
the rich phenomenology described here, and they can be used to test some 
of our quantitative predictions, concerning the values of the exponents 
and the existence of various universality classes. Liquid-crystals
\cite{LQ}, molecular-beam epitaxial systems, as $GaAS$ \cite{GAAS}
claimed to grow following KPZ scaling, or materials exhibiting
Stranski-Krastanov instabilities \cite{SK}, appear to be good candidates
that are at least worth investigating in this context.
Indeed, it is rather exciting to think that non-equilibrium 
complete wetting exponents are measurable. This would be a way of
measuring the multiplicative noise critical exponents, and brings new
hope of measuring directed percolation exponents in real systems \cite{DPexperiments}.

\vspace{0.5cm}
{\bf Acknowledgments} {\small We acknowledge financial support from
MEyC-FEDER, project FIS2005-00791, and from Junta de Andaluc{\'\i}a as
group FQM-165. The ``Acci\'on Integrada hispano-portuguesa''
HP2003-0028 is also acknowledged.}

\appendix*
\section{BRIEF REVIEW OF EQUILIBRIUM WETTING}

The action associated with Eq. (\ref{EWwall})
\cite{GF,ZJ} (setting $c=0$) is
\begin{equation}
{\cal S}(h,\tilde h) = \int d^d x dt \left\{ \tilde h^2 -\tilde
h\left[\partial_t h - \nabla^2 h -a - b h^{-p-1}\right]\right\},
\end{equation}
where $\tilde h$ denotes, as usual, the response field \cite{GF,ZJ}.
If one assumes first that the interaction term is the dominant one,
from na\"ive dimensional analysis, imposing $b$ to be dimensionless at
the upper critical dimension, and equating the dimensions of the
time-derivative and the potential terms, one obtains
$[h]_{MF}=L^{2/(p+2)}$ and consequently, within mean-field,
$\theta_h=1/(p+2)$ since time scales na\"ively as $L^{2}$. The
exponent values $\beta_h=1/(p+1)$ and $\nu= (p+2)/(2p+2)$ are then
obtained by matching $[a]=[h]_{MF}^{-p-1}$ and by identifying $L$ with
the characteristic correlation length, respectively.

On the other hand, when fluctuations (i.e. the noise term) dominate,
we require the noise amplitude to be dimensionless at the upper critical
dimension, which leads to $[\tilde h]_{FL}=L^{(2+d)/2}$, and
therefore $[h \tilde h] = L^{-d}$, $[h]_{FL}=L^{(2-d)/2}$. From
this, proceeding as before, $\theta_h=(2-d)/4, \nu=2/(d+2)$, and
$\beta_h = (2-d)/(d+2)$. These results (which may be obtained using a
number of different procedures \cite{Zia,LipoFisher,Huse,FisherFisher,Lipowsky85,reviews_wetting})
are exact as long as $h$ and $\tilde h$ do not have anomalous dimensions,
which has been shown to be the case \cite{LipoFisher}.

The upper critical dimension is defined by $[h]_{MF}=[h]_{FL}$, which
yields $d_c(p)=2p/(2+p)$. Note that for $d>d_c(p)$ the critical
exponents depend on the details of the interaction (i.e. on $p$)
whilst for $d<d_c(p)$, they depend only on $d$.  In
particular, in one dimensional systems, $p=2$ marks the transition
between a {\it mean-field regime} and a {\it fluctuation regime}:

{\bf i)} If $p < 2$ mean field theory is valid, and consequently
$\theta_h=1/(p+2)$, $\beta_h=1/(p+1)$, $z=2$, and $\nu=1/2$.

{\bf ii)} For $p \geq 2$, the substrate interaction decays fast enough for
the fluctuations to take over and the exponents become $p$-independent:
$\theta_h=1/4$, $\beta_h=1/3$, $z=2$, and $\nu=2/3$.  

Note that at the limiting value $p=2$ the exponents change
continuously from the mean field to the fluctuation regime. It is also 
remarkable that the fluctuation regime exponents coincide with those of short-ranged 
equilibrium wetting (characterized by exponential bounding
potentials \cite{reviews_wetting}). 

Until now we have considered the scaling properties of $\langle h
\rangle$, but as was mentioned earlier the number of dry sites 
or contact points between the interface and the substrate, measured  
by $\langle\exp(-h) \rangle$, is known to exhibit interesting 
scaling behavior in wetting problems \cite{surface}. 

{\bf i)} For $p < 2$ simple mean field scaling holds, and the
$h$-distribution is a Gaussian detaching from the wall at a speed
controlled by its mean value. As the interface is well described 
by its average position, it is expected that
\begin{equation}
\langle e^{-h} \rangle \sim e^{-\langle h \rangle} 
\sim e^{- A t^{1/(p+2)}},
\end{equation}
yielding a {\it stretched exponential} decay.

{\bf ii)} For $p > 2$, $\langle a + b \exp (-h) \rangle =0 $ holds 
in the stationary state, and therefore $\langle n \rangle \propto a$; 
using simple scaling, $[\exp(-h)] = [a] = [\partial_t h] \sim
t^{1/4-1}$, giving $\langle \exp(-h) \rangle \sim t^{-3/4} $. This
result can be derived in a number of ways, including explicit
calculations for discrete models in this class
\cite{Ginelli1}, and remains valid for long-ranged potentials in the 
fluctuation regime. 
Note the difference between this
fluctuation-induced power-law behavior and the previously reported
stretched exponential behavior in the mean-field regime.

For attractive walls, $b<0$, a positive value of $c$ is required to 
ensure the impenetrability of the substrate (see Fig. \ref{fig1}). In this 
case it is easy to argue that the interface jumps discontinuously 
from a bound state (for $a<0$), localized at the minimum of $V(h)$, 
to an unbound state (for $a>0$) through a first-order phase transition.
Clearly, in terms of the contact points, $\langle \exp(-h) \rangle $, 
the transition is also discontinuous.


\begin{thebibliography}{99}
\bibitem{2d}
Note that for higher dimensional interfaces, the mean field regime
dominates for any $p$, $1/(p+2) > 0$, as the free fluctuations are flat, 
being characterized by a zero roughness exponent and a logarithmic growth
in $d=2$.
\bibitem{MBE} Another possibility (presently under study) is to analyze 
molecular beam epitaxy (MBE) Langevin equations (known to represent a
broad family of non-equilibrium interfacial growth \cite{HZ}) in the
presence of bounding potentials.
\bibitem{transient} The true straight-line boundary may be easily
mistaken for a temperature-dependent curve. This is due to a roughness
suppression effect that effectively switches off the KPZ nonlinear
term when the interfaces are initially placed close to the substrate
and within the potential well. In this case, a transient renormalized
$\lambda_R(b)$ may be observed that leads to an apparent
temperature-dependent transition $a+\lambda_R(b) \langle (\nabla h)^2
\rangle =0$.
\end{thebibliography}
\end{document}